\documentclass[doublecol]{epl2}
\usepackage{graphicx}
\usepackage{amsmath}
\usepackage{amsfonts}
\usepackage{color}
\usepackage{bm}
\newcommand{\beq}{\begin{equation}}
\newcommand{\eeq}{  \end{equation}}
\newcommand{\bea}{\begin{eqnarray}}
\newcommand{\eea}{  \end{eqnarray}}
\newcommand{\bit}{\begin{itemize}}
\newcommand{\eit}{  \end{itemize}}

\newcommand{\gsim}{\mathrel{\hbox{\rlap{\lower.55ex 
\hbox{$\sim$}} \kern-.3em \raise.5ex \hbox{$>$}}}}

\begin{document}

\title{Heisenberg Uncertainty Relation for Coarse-grained Observables}

\author{{\L}ukasz ~Rudnicki\inst{1}\thanks{E-mail: \email{rudnicki@cft.edu.pl}}, Stephen P. ~Walborn\inst{2} \and Fabricio Toscano\inst{2}}
\shortauthor{{\L}ukasz ~Rudnicki \etal}
\institute{                    
  \inst{1} Center for Theoretical Physics, Polish Academy of Sciences, Aleja
Lotnik{\'o}w 32/46, PL-02-668 Warsaw, Poland\\
  \inst{2} Instituto de F{\'{i}}sica, Universidade Federal do Rio de Janeiro,
Caixa Postal 68528, Rio de Janeiro, RJ 21941-972, Brazil
}

\abstract{
We ask which is the best strategy
to reveal uncertainty relations between complementary observables of a continuous variable system for coarse-grained measurements.  This leads to the derivation of new uncertainty relations for coarse-grained measurements
that are always valid, even for detectors with low precision.  
These relations should be particularly relevant in experimental demonstrations
of squeezing in quantum optics, quantum state reconstruction, and the development of trustworthy  entanglement criteria.}

\pacs{03.65.Wj}{State reconstruction, quantum tomography}
\pacs{89.70.Cf}{Entropy and other measures of information}
\pacs{03.65.Ca}{Quantum mechanics -- Formalism}
\pacs{03.67.Mn}{Entanglement measures, witnesses, and other characterizations}
\maketitle
\date{\today}

\section{Introduction}

In Quantum Mechanics, mathematical inequalities that originate from the fact that measured quantities are directly associated to non-commuting operators are generically called uncertainty relations.
The best known of these inequalities is the Heisenberg uncertainty relation (HUR), which
sets a bound on the product of the variances of  two 
complementary observables \cite{Heisenberg1927}.
In quantum mechanical systems of continuous variables (CV), the Hilbert space is spanned by the
eigenstates $|x\rangle$ and $|p\rangle$ of canonical operators $\hat x$ and $\hat p$ ($[\hat x, \hat p]=i\hbar$),
that without loss of generality we can call the position and the momentum.  For these variables, the HUR is
\beq
\label{Heisenberg-uncertainty-rel}
\sigma_{x,\rho}^2  \sigma_{p,\tilde\rho}^2 \geq \frac{1}{4}|\langle[\hat x,\hat p]\rangle|^2=\frac{\hbar^2}{4}, \eeq
where $\sigma_{x,\rho}^2 \equiv
\langle\hat x^2\rangle-\langle\hat x\rangle^2$ and 
 $\sigma_{p,\tilde\rho}^2 \equiv
\langle\hat p^2\rangle-\langle\hat p\rangle^2$ are the variances of  $ x$ and $ p$ measurements on some (in general mixed) quantum state $\hat\rho$.   
 The variances in (\ref{Heisenberg-uncertainty-rel}) are obtained from the probability densities: 
 \beq
 \label{disttrue}
\rho(x)= \langle x|\hat \rho|x\rangle\;\; \mbox{and}\;\;
 \tilde \rho(p)= \langle p|\hat \rho|p\rangle.
 \eeq
 {Except for the fact that $\rho(x)$ and $\tilde\rho(p)$  are marginal distributions of the Wigner function
 of the quantum state}, we note that these densities have all the properties of classical probability density functions (PDFs) of continuous variables.  
In addition to the uncertainty relation for the variances, there is the Bialynicki-Birula-Mycielski uncertainty relation (BBM) involving the Shannon entropies of these PDFs \cite{Bialinicki-Birula-old},
\beq
\label{old-Bialinicki-result}
h[\rho]+h[\tilde \rho]\geq \ln\left(\pi e \hbar\right),
\eeq
where the Shannon entropy is defined as:
$h[f]\equiv
-\int_{\mathbb{R}}
\; dz\, f\left(z\right)\;\ln\left[f(z)\right]$.
Using  the reversed logarithmic Sobolev inequality \cite{sobole} for PDFs:
$\ln\left[2\pi e\,\sigma_{z,f}^{2}\right]\geq 2 h[f]$,
one can compress the inequalities (\ref{Heisenberg-uncertainty-rel}) 
and (\ref{old-Bialinicki-result}) into
\beq
\label{complete-ineq}
\ln\left(2\pi e \sigma_{x,\rho} \sigma_{p,\tilde\rho}\right)\geq 
h[\rho]+h[\tilde \rho]\geq \ln\left(\pi e \hbar\right).
\eeq
\par
Uncertainty relations are intrinsic features of quantum states \cite{Dodonov1989} and 
their fullfillment can be considered as a sufficient criterion for a quantum mechanically permissible state 
\cite{Nha2008}. 
An experimental test of an uncertainty relation is not only a way to establish the  quantum nature
of a physical system but is also a way to characterize or identify salient quantum 
features. Estimates of uncertainty are used to characterize nonclassical states of
radiation fields \cite{slusher85}, and are useful to construct  entanglement
criteria for the whole class of negative partial-transpose states 
\cite{Nha2008,walborn09,saboia11}.  Security in certain quantum cryptography
protocols rely on the violation of uncertainty criteria by post-selected ensembles
\cite{reid00,grosshans04}. Thus, the evaluation of uncertainty relations from experimental data is important for both fundamental studies of quantum physics as well as applications. 
\par
The uncertainty relations contained in inequalities \eqref{complete-ineq} concern CV, whereas experiments are always performed with finite precision.   This imposes sampling widths $\Delta$ and $\delta$ on the $x$ and $p$ measurements,  respectively. { The position and momentum spaces should then be divided into bins (labelled by $k$ and $l$) describing  the discrete sampling \cite{ibb0,sen}, where the corresponding bin widths are  $\Delta$ and $\delta$.} Thus, what one obtains experimentally are discrete probability distributions $\{r_k^{\Delta}\}$ and $\{s_l^{\delta}\}$ corresponding to the 
different coarse-grained measurements of $x$ and $p$, respectively.
In the limit that $\Delta,\delta \longrightarrow 0$, these coarse-grained distributions reproduce the probability densities \eqref{disttrue}.  
The problem that arises is how to evaluate the uncertainty relations with the coarse-grained quantities in the general case, so that even when $\Delta$ and $\delta$ are not sufficiently small one still obtains a reliable estimate of an uncertainty relation.  As an example, consider the { extreme} case where $\Delta$ is larger than the support of $\rho(x)$.  Then it is possible that only one $r_k^{\Delta}$ is non-zero.  In this case the uncertainty in the corresponding discrete distribution is zero (we know in which bin the particle is localized), and assumption that $\{r_k^{\Delta}\}$ and $\{s_l^{\delta}\}$  are accurate representations of the quantum state can result in a \textit{false violation} of the HUR \eqref{Heisenberg-uncertainty-rel} or BBM \eqref{old-Bialinicki-result}.  
\par
In this work, we show that there exist two reliable strategies to address this problem, where we restrict ourselves to  the case  of  the uncertainty relations  compressed in the inequalities (\ref{complete-ineq}).   The first one uses the uncertainty estimates corresponding to the discrete distributions  $\{r_k^{\Delta}\}$ and $\{s_l^{\delta}\}$ directly.  
It is then necessary to translate the uncertainty relations in the
 inequalities (\ref{complete-ineq}) for these discrete quantities.
This involves finding appropriate lower bounds in the uncertainty relations which shall depend on the sampling windows widths $\Delta$ and $\delta$. This problem can be avoided by adopting a second strategy.
In this approach, the discrete probability distributions   $\{r_k^{\Delta}\}$ and $\{s_l^{\delta}\}$ are used as the relative frequencies of histograms that approximate the probability densities $\rho(x)$ and $\tilde\rho(p)$.  These histograms constitute coarse-grained PDFs of the continuous variables.
We show that, surprisingly, the uncertainty estimates of these coarse grained PDFs always satisfy the relevant uncertainty relations contained in (\ref{complete-ineq}) and thus can be employed as reliable estimates of uncertainty.  { From an experimental point of view, this second strategy does not require any extra effort with respect to the first, and has the advantages that, in addition to being quite intuitive, explicit expressions for the uncertainty lower bounds can be obtained.} 
\par
Let us start by defining the hermitian operators that represent finite precision position and momentum measurements: 
\begin{subequations}
\beq
\hat{x}_{\Delta}=\sum_{k}x_{k}\int_{(k-1/2)\Delta}^{(k+1/2)\Delta}dx\, 
\left|x\right\rangle \left\langle x\right|,\label{pcoar}
\eeq
\beq
\hat{p}_{\delta}=\sum_{l}p_{l}\int_{(l-1/2)\delta}^{(l+1/2)\delta}dp\,
\left|p\right\rangle \left\langle p\right|,\label{mcoar}
\eeq
\end{subequations}
where we choose $x_k\equiv k\Delta$ and $p_l=l\delta$ as the centers of the sampling windows. From the spectral representations of these operators we have {the} limits $\lim_{\Delta\rightarrow 0}\;\hat{x}_{\Delta}=\hat{x}$ and $\lim_{\delta\rightarrow 0}\;\hat{p}_{\delta}=\hat{p}$, { as expected}.
{ Repeated measurements over identically prepared systems allow for the construction of the probabilities} $r_k^{\Delta}$ and $s_l^{\delta}$
to obtain the results $x_k$ and  $p_l$, respectively.
By definition the probabilities $r_k^{\Delta}$ and $s_l^{\delta}$ 
should verify the normalization condition, {\it i.e.},
\beq
\sum_k r_k^{\Delta}=1 \;\;\;\mbox{and}\;\;\;\sum_l s_l^{\delta}=1.
\eeq 
We assume that the position and momentum measurements were performed with sufficiently large statistics so these probabilities are expected to be very close to the theoretical values,
\begin{subequations}
\beq
\label{rk}
r_{k}^{\Delta}=\int_{\left(k-1/2\right)\Delta}^{\left(k+1/2\right)\Delta}dx\,\rho\left(x\right)
=\int_{\mathbb{R}}dx\, {\mathbb I}_{\Delta}\left(x,x_{k}\right)\rho\left(x\right),
\eeq
and
\beq
\label{sl}
s_{l}^{\delta}=\int_{\left(l-1/2\right)\delta}^{\left(l+1/2\right)\delta}dp\,\tilde{\rho}\left(p\right)=\int_{\mathbb{R}}\;dp\; {\mathbb I}_{\delta}\left(p,p_{l}\right)
\tilde{\rho}\left(p\right),
\eeq
\end{subequations}
where we have introduced the rectangle function
\beq
{\mathbb I}_{\eta}\left(z,z_{j}\right)=\begin{cases}
1 & \textrm{ for }z\in\left[\left(j-\frac{1}{2}\right)\eta,\left(j+\frac{1}{2}\right)\eta\right]\\
0 & \textrm{ elsewhere}\end{cases}
\label{rect}.
\eeq

The first strategy that can be used to check the HUR
(\ref{Heisenberg-uncertainty-rel}) consists in using the discrete variances:
\begin{subequations}
\beq
\label{def-discrete-position-variance}
\sigma_{x_{\Delta}}^2\equiv\langle \hat{x}_{\Delta}^2\rangle-\langle\hat{x}_{\Delta}\rangle^2=
\sum_{k}x_k^2r_k^{\Delta}-\left(\sum_k x_k r_k^{\Delta}\right)^2,
\eeq
and
\beq
\label{def-discrete-momentum-variance}
\sigma_{p_{\delta}}^2\equiv\langle\hat{p}_{\delta}^2\rangle-\langle\hat{p}_{\delta}\rangle^2=
\sum_{l}p_l^2s_l^{\delta}-\left(\sum_l p_l s_l^{\delta}\right)^2,
\eeq
\end{subequations}
that correspond to the operators (\ref{pcoar}) and (\ref{mcoar}).  These variances describe the uncertainty of position and momentum { (in units of bin-width)} that can be inferred from the discrete 
probability distributions $\{r_k^{\Delta}\}$ and $\{s_l^{\delta}\}$ respectively.
In the case where the widths of the detectors $\Delta$ and $\delta$ are sufficiently small,
these discrete variances are good approximations of the variances $\sigma^2_{x,\rho}$ and 
$\sigma^2_{p,\tilde \rho}$ in the inequality (\ref{Heisenberg-uncertainty-rel}).
However, it is easy to recognize that as these widths start to grow the inferred variances 
$\sigma_{x_{\Delta}}^2$  and $\sigma_{p_{\delta}}^2$ start to underestimate the true variances
$\sigma^2_{x,\rho}$ and  $\sigma^2_{p,\tilde \rho}$ respectively. In fact we have the limits,
$\lim_{\Delta\rightarrow \infty}\;\sigma_{x_{\Delta}}^2=0$ and $\lim_{\delta\rightarrow \infty}\;\sigma_{p_{\delta}}^2=0$.
For the product in general we have $0\leq\sigma_{x_{\Delta}}^2 \sigma^2_{p_{\delta}}\leq \sigma_{x,\rho}^2\sigma_{p,\tilde\rho}^2$, so for any normalizable quantum state there always exist some values of $\Delta$ and $\delta$ such that $\sigma_{x_{\Delta}}^2 \sigma^2_{p_{\delta}}<\hbar^2/4$, which is a false violation of the HUR. Thus, when the coarse-graining 
windows are not sufficiently small, the variances $\sigma_{x_{\Delta}}^2$ and $\sigma^2_{p_{\delta}}$ are not reliable estimates of the true variances, and cannot be used to
check the HUR (\ref{Heisenberg-uncertainty-rel}). 
It is necessary to find a $\Delta$ and $\delta$ dependent uncertainty relation  for the variances of the 
discrete probability distributions $\{r_k^{\Delta}\}$ and $\{s_l^{\delta}\}$. It should be of the general form $f\left(\sigma_{x_{\Delta}}, \sigma_{p_{\delta}},\Delta,\delta \right)\geq \hbar^2/4$. { To our knowledge, the} optimal choice of the $f$ function is not known, but we should have  $\lim_{\Delta\rightarrow 0}\lim_{\delta \rightarrow 0} f\left(\sigma_{x_{\Delta}}, \sigma_{p_{\delta}},\Delta,\delta \right)=\sigma_{x,\rho}^2\sigma_{p,\tilde \rho}^2$. The existence of such nontrivial uncertainty relation stems from the fact that the discrete probability distributions $\{r_k^{\Delta}\}$ and $\{s_l^{\delta}\}$ are obtained from complementary probability distributions $\rho (x)$ and $\tilde{\rho} (p)$ (see Eq. (\ref{rk}) and (\ref{sl})).  We are unaware of any results concerning such explicit lower bound (cf. \cite{Busch}).
\par
\begin{figure}
\begin{center}
\includegraphics[width=7cm]{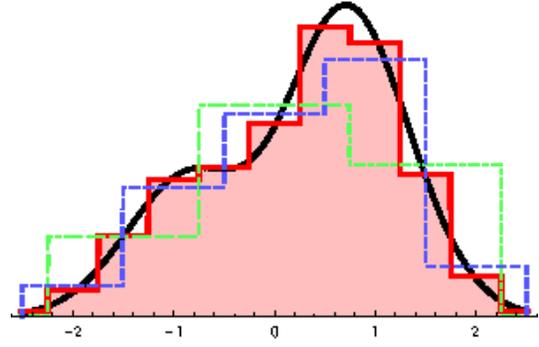}
\end{center}
 \caption{(color online). Examples of binned approximations to a continuous probability distribution for different detectors’ widths: $\eta =3/2$  (green, dashed-dotted); $\eta = 1$
(blue, dashed); $\eta = 1/2$  (red, filled).}
\label{fig:1}
 \end{figure}

  
We would like to use a slightly different strategy to verify the uncertainty relations associated with coarse-grained measurements.   Taking into account the same coarse-grained measurement results, we use the following  PDFs: 
\begin{subequations}
\beq
\label{def-w-Delta}
w_{\Delta}\left(x\right)=\sum_{k}r_{k}^{\Delta}D_{\Delta}\left(x,x_{k}\right),
\eeq
and 
\beq
\label{def-tilde-w-delta}
\tilde{w}_{\delta}\left(p\right)=\sum_{l}s_{l}^{\delta}D_{\delta}\left(p,p_{l}\right).
\eeq
\end{subequations}
In (\ref{def-w-Delta}) and (\ref{def-tilde-w-delta}) the function
$D_{\eta}\left(z,z_{j}\right)={\mathbb I}_{\eta}\left(z,z_{j}\right)/\eta$ is 
a normalized version of the rectangle function (\ref{rect}), that converges to the
Dirac delta function $\lim_{\eta\rightarrow0}=D_{\eta}(z,z_j)=
\delta(z-z_j)$.  Each rectangle $D_{\eta}\left(z,z_{j}\right)$ in the expansion is weighted by the $j$th value of the associated discrete probability.  
Thus,  the $w$ and $\tilde w$ PDFs, that were also used before in \cite{raymer}, serve as reconstructions of the actual PDFs \eqref{disttrue}
from the measurements with finite resolution detectors. We emphasize that 
$\Delta$ and $\delta$ are the widths of the resolution of the detectors and 
in principle are independent.
As an example, Fig. \ref{fig:1} illustrates several functions $w_{\Delta}(x)$ as reconstructions to
a given probability density $\rho(x)$, the accuracy of which depends of course on the width $\Delta$ of the sampling windows.  
In the limit where the 
resolution widths of the detectors go to zero, we have
$\lim_{\Delta\rightarrow 0} w_{\Delta}(x)= 
\rho(x)$ and 
$\lim_{\Delta\rightarrow 0} \tilde w_{\delta}(p)=
\tilde\rho(p)$
\footnote{Here we use that 
$\lim_{\eta\rightarrow 0} {\mathbb I}_{\eta}\left(z,z_{j}\right)/\eta=\delta(z-z_j)$
and $\lim_{\eta\rightarrow 0}\sum_j \eta=\int dz_j$. }.
However, for finite sampling widths, clearly the coarse-grained probability density $w_{\Delta}(x)$ and $\tilde w_{\delta}(p)$ are not necessarily close approximations to the actual PDFs associated to the quantum state in question via Eqs \eqref{disttrue}.  Consequently, there is no reason to expect a priori that the uncertainty estimates (variance or entropic) obtained from these densities should necessarily obey the uncertainty relations compressed in Eq.(\ref{complete-ineq}). Surprisingly, we will now show that \eqref{def-w-Delta} and \eqref{def-tilde-w-delta}, though coarse-grained estimates of the actual PDFs, do indeed obey the Heisenberg and BBM uncertainty relations.  
\par
First we note that the probability distributions $w_{\Delta}(x)$ and $\tilde w_{\delta}(p)$ necessarily carry less information (more uncertainty) than $\rho (x)$ and $\tilde{\rho} (x)$, and we have the following relations:
\beq
\label{ineq-hw-hrho}
h[w_{\Delta}]\geq h[\rho]\;\;\mbox{and}\;\;
h[\tilde w_{\delta}]\geq h[\tilde \rho].
\eeq
To prove (\ref{ineq-hw-hrho}) explicitly we shall notice that  
\beq
\label{rel-entropy1}
h[w_{\Delta}]=
H[r_k^{\Delta}]+\ln({\Delta})\;\;
\mbox{and}\;\;
h[\tilde w_{\delta}]=
H[s_l^{\delta}]+\ln({\delta}),  
\eeq
and 
\beq
\label{Jensen}
H[r_k^{\Delta}]\geq h[\rho] -\ln({\Delta})\;\;\mbox{and}\;\;
H[s_l^{\delta}]\geq h[\tilde \rho] -\ln({\delta}),
\eeq
that can easily be proved using Jensen's inequality \cite{Cover}. Here, the discrete Shannon entropies are:
\beq
\label{discrete-entropies}
H[r_k^{\Delta}]\equiv-\sum_k r_k^{\Delta}\ln r_k^{\Delta}\;\;\;
\mbox{;}\;\;\; 
H[s_l^{\delta}]\equiv-\sum_l s_l^{\delta}\ln s_l^{\delta}.
\eeq
Now we shall use the same reversed logarithmic Sobolev inequality \cite{sobole} for the probability distributions (\ref{def-w-Delta}) and (\ref{def-tilde-w-delta}) together with  (\ref{ineq-hw-hrho}) and (\ref{old-Bialinicki-result}), and obtain
\beq
\label{ineq-hwx-hwp}
\ln\left(2\pi e \sigma_{x,w_{\Delta}} \sigma_{p,\tilde w_{\delta}}\right)\geq
h[w_{\Delta}]+h[\tilde w_{\delta}] \geq \ln\left(\pi e \hbar\right).
\eeq
The result (\ref{ineq-hwx-hwp}) is valid in spite of the fact that $w_{\Delta}(x)$ and $\tilde w_{\delta}(p)$ are not marginal probability distributions calculated from the Wigner function of a quantum state $\hat \rho$.
Therefore, we can faithfully evaluate the uncertainty relations compressed in 
Eq.(\ref{ineq-hwx-hwp}) using only the variances $\sigma_{x,w_{\Delta}}^2$,  $\sigma_{p,\tilde w_{\delta}}^2$ or Shannon entropies $h[w_{\Delta}], h[\tilde w_{\delta}]$  of the 
probability distributions $w_{\Delta}(x)$, $\tilde w_{\delta}(p)$.
\par
From  Eqs.(\ref{rel-entropy1} we note that the entropic uncertainty relation in Eq.(\ref{ineq-hwx-hwp}) corresponds to the 
previously known uncertainty relation for discrete entropies \cite{ibb0}: 
$H[r_k^{\Delta}]+H[s_l^{\delta}]\geq  \ln\left(\pi e \hbar\right)-
\ln\left(\Delta\delta\right)$. For $\Delta\delta\geq \pi e \hbar$ this is trivially satisfied
because the discrete entropies are always positive.  
However, Eq.(\ref{ineq-hwx-hwp}) also establishes that the variances of the coarse-grained distributions satisfy the HUR,
\beq
\sigma_{x,w_{\Delta}}^2 \sigma_{p,\tilde w_{\delta}}^2\geq 
\frac{\hbar^2}{4}.
\label{HURnew}
\eeq
A similar uncertainty relation was derived in \cite{raymer}, 
however it had two important differences comparing with the 
uncertainty relation in Eq. \eqref{HURnew}. The first one, is that the
uncertainty relation in \cite{raymer} is a direct consequence of 
the joint measurements of two noncommuting variables whose unavoidable 
imprecision introduce additional uncertainties beyond those in the usual HUR 
uncertainty relation. Thus, the joint measurement 
necessarily relates the widths of the resolution of position and momentum
measurement as $\delta=2\pi\hbar/\Delta $. The new HUR in Eq. \eqref{HURnew}
is not for joint measurement of position and momentum but it is a statistical 
consequence of finite resolution measurement of two noncommuting observables
whose coarse-grained widths $\Delta$ and $\delta$ are independent.
The second difference is that the uncertainty relation in  \cite{raymer} shall be
only valid for sufficiently small values of coarse graining where 
essentially we have $\sigma_{x,w_{\Delta}}^2\sim \sigma_{x,\rho}^2 $ and
$\sigma_{p,\tilde w_{\delta}}^2 \sim \sigma_{p,\tilde \rho}^2 $. In contrast, the HUR 
uncertainty relation in Eq. \eqref{HURnew} is for arbitrary  values of 
$\Delta$ and $\delta$.

The result in Eq. \eqref{HURnew} could be obtained trivially if, similarly to (\ref{ineq-hw-hrho}), $\sigma_{x,w_{\Delta}}^2 \geq 
\sigma_{x,\rho}^2$ and $\sigma_{p,\tilde w_{\delta}}^2\geq \sigma_{p,\tilde\rho}^2$.
In order to convince ourselves that this is not the case and that inequality (\ref{HURnew}) is more sophisticated than one might think, let us briefly investigate the following state in position space ($\kappa\in \mathbb R$):
\begin{equation}
\rho_{\kappa}\left(x\right)=\begin{cases}
\mathcal{N}\exp\left(-\kappa x^{2}\right) & x\in\left[-\Delta/2,\Delta/2\right]\\
0 & \textrm{elsewhere}
\end{cases}.
\end{equation}
The normalization constant reads $\mathcal{N}=\sqrt{\pi/\kappa} \textrm{Erf}\left(\Delta\sqrt{\kappa}/2\right)$. Since this state is localized in the interval $\left[-\Delta/2,\Delta/2\right]$, we have $\sigma_{x_{\Delta}}^{2}=0$ and $\sigma_{x,w_\Delta}^{2}=\Delta^2/12$.
We shall also calculate explicitly
\begin{equation}
\sigma_{x,\rho}^{2}\left(\kappa\right)=\frac{1}{2\kappa}-\frac{\Delta}{2\sqrt{\kappa\pi}}\frac{\exp\left(-\kappa\Delta^{2}/4\right)}{\textrm{Erf}\left(\Delta\sqrt{\kappa}/2\right)}.
\end{equation}
One finds that:
\begin{equation}
\sigma_{x,w_\Delta}^{2}>\sigma_{x,\rho}^{2}\left(\kappa>0\right),\qquad \sigma_{x,w_\Delta}^{2}<\sigma_{x,\rho}^{2}\left(\kappa<0\right),
\end{equation}
and thus inequality \eqref{HURnew} is not a trivial extension of the usual HUR \eqref{Heisenberg-uncertainty-rel}. 
\par
The new inequality (\ref{HURnew}) provides an uncertainty relation directly for the
discrete variances \eqref{def-discrete-position-variance} and \eqref{def-discrete-momentum-variance}, since it is straightforward to show that \cite{raymer}:
\beq
\label{relposVariance}
\sigma_{x,w_{\Delta}}^{2}=
\sigma_{x_{\Delta}}^2+\sigma^2_{\Delta}\;\;
\mbox{and}\;\;
\sigma_{p,\tilde w_{\delta}}^{2}=
\sigma_{p_{\delta}}^2 + \sigma^2_{\delta}, 
\eeq
where $\sigma^2_{\eta}\equiv\eta^2/12$ is the variance of the normalized rectangle function 
$D_{\eta}(z,z_j)$.  

\begin{figure}
\begin{centering}
\includegraphics[width=6cm]{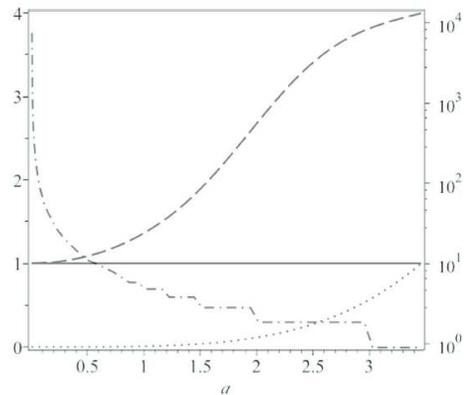}
\par\end{centering}
\caption{The scale on the left is for the dashed and dotted curves, $\sigma_{x,\omega_{\Delta}}^2\sigma_{p,\tilde\omega_{\delta}}^2/(\hbar^2/4)$ vs $a$ and $\sigma_{\Delta}^2\sigma_{\delta}^2/(\hbar^2/4)$ vs $a$ respectively ($a=\Delta/\sigma_{x,\rho}=\delta/\sigma_{p,\tilde\rho}$), for a minimum uncertainty state (see the text for details). 
When the dotted curve passes the straight solid line at $y=1$ (i.e. when
$a=\sqrt{12}$) our HUR in Eq.(\ref{uncer2}) is satisfied trivially.
The logarithm scale on right is for the dash-dotted curve that represents the  number of measurements (bins) of discrete position and momentum to reconstruct the PDFs in the intervals
$-6\sigma_{x,\rho}\leq x_k \leq 6 \sigma_{x,\rho}$ and $-6\sigma_{p,\tilde\rho}\leq p_l \leq 6 \sigma_{p,\tilde\rho}$ respectively. }
\label{fig2}
\end{figure}
Then we can write explicitly
\beq
\sigma_{x_{\Delta}}^2  \sigma_{p_{\delta}}^2 + (\sigma^2_{\Delta}  \sigma_{p_{\delta}}^2 + \sigma^2_{\delta}\sigma_{x_{\Delta}}^2)
+\sigma^2_{\Delta} \sigma^2_{\delta} 
\geq
\frac{\hbar^2}{4} .
\label{uncer2}\eeq 
\par
This is the generalization of the HUR { for 
coarse-grained measurements}, and in the limit $\Delta, \delta \longrightarrow 0$
we recover the relation (\ref{Heisenberg-uncertainty-rel}). 
We
recall that the uncertainty
relation (\ref{uncer2}) is valid independently of the coarse graining
widths $\Delta$ and $\delta$.
Validity and significance of our uncertainty relation becomes more clear if we set the phase space origin 
at the center of the bins that contain $\langle \hat x_{\Delta}\rangle$ and $\langle \hat p_{\delta}\rangle$ ({\it i.e.} the central bins).  From Eqs.  \eqref{def-discrete-position-variance} and \eqref{def-discrete-momentum-variance} we can interpret the two contributions to the variances defined in Eqs.(\ref{relposVariance})
in the following way.   The first contribution is given by the discrete variances $\sigma^2_{x_{\Delta}}$ and $\sigma^2_{p_{\delta}}$ corresponding to the coarse grained measurements $x_k=k\Delta$ and $p_l=l\delta$ outside the central bins, since the central bin has no contribution to the discrete variance in this case.  The other contributions,  $\sigma^2_{\Delta}$  and $\sigma^2_{\delta}$, can be interpreted as the  variance of the  central bins  of the histograms. 
For increasing values of coarse graining, the contribution from each central bin grows and the contribution from discrete measurements outside the central bins decreases. When the phase space area of the central bins reaches the value  of the minimum uncertainty area in phase space, {\it i.e.}
when $\sigma^2_{\Delta}\sigma_{\delta}^2=\hbar^2/4$, the generalized HUR relation in Eq.(\ref{uncer2}) is satisfied trivially. This is illustrated 
in Fig. \ref{fig2}, where we apply our HUR in Eq.(\ref{uncer2}) to a minimum uncertainty squeezed vacuum state ($\sigma^2_{\rho}\sigma^2_{\tilde\rho}=\hbar^2/4$), whose squeezing/anti-squeezing directions are aligned with the $x$ and $p$ axis. 
In this example our HUR is satisfied trivially only when $\Delta\geq \sqrt{12}\,\sigma_{x,\rho}$
and $\delta \geq\sqrt{12}\,\sigma_{p,\tilde\rho}$.  However, well before that,
and even for great value of the detectors widths $\Delta$ and $\delta$, 
we can verify the uncertainty principle non-trivially with a very low number of measurements (bins).  For example when $\Delta=\sigma_{x,\rho}$ and
$\delta=\sigma_{p,\tilde\rho}$ we only need on the order of $10$ measurements. 

\section{Conclusion}  We have shown that care must be taken when evaluating uncertainty relations using experimental data.  In particular, coarse-grained measurements can lead to a false violation of uncertainty relations.  
Here we derive new uncertainty relations that are valid for any size detector or sampling window.  These new relations are always verified by any physical state, and should be relevant in fundamental investigations of quantum physics as well as applications such as quantum cryptography.

 \begin{acknowledgements}
 We acknowledge financial support from the Brazilian funding agencies
CNPq and FAPERJ, and the
INCT - Informa{\c c}{\~a}o Qu{\^a}ntica.
This research was also supported by the grant number N N202 174039 from the Polish Ministry
of Science and Higher Education for the years 2010\textendash{}2012. SPW thanks F. Brito for helpful discussions.
\end{acknowledgements}

\end{document}